\documentclass[aps,prd,superscriptaddress,showpacs,amsfonts,twocolumn]{revtex4}

\begin{document}


\title{Conservative special relativistic radiative transfer for multidimensional astrophysical simulations: motivation and elaboration}



\author{Christian Y. Cardall}
\email[]{cardallcy@ornl.gov}
\affiliation{Physics Division, Oak Ridge National Laboratory, Oak Ridge,
 	TN 37831-6354}
\affiliation{Department of Physics and Astronomy, University of Tennessee,
	Knoxville, TN 37996-1200} 

\author{Eric J. Lentz}
\email[]{lentzej@ornl.gov}
\affiliation{Physics Division, Oak Ridge National Laboratory, Oak Ridge,
 	TN 37831-6354}
\affiliation{Department of Physics and Astronomy, University of Tennessee,
	Knoxville, TN 37996-1200} 
\affiliation{Joint Institute for Heavy Ion Research, Oak Ridge National
	Laboratory, Oak Ridge, TN 37831-6374}

\author{Anthony Mezzacappa}
\email[]{mezzacappaa@ornl.gov}
\affiliation{Physics Division, Oak Ridge National Laboratory, Oak Ridge,
 	TN 37831-6354}


\date{\today}

\def\InvisibleSpace{}
\def\InvisibleComma{}
\def\Mvariable{}
\def\Mfunction{}
\def\dispSFNumberedEquationmath{}
\def\inlineSFinmath{}
\def\multsp{}
\def\IndentingNewLine{}

\def\Tilde{\sim}
\def\ScriptCapitalL{{\cal L}}
\def\ScriptCapitalN{{\cal N}}
\def\ScriptCapitalT{{\cal T}}
\def\ScriptCapitalO{{\cal O}}
\def\ScriptCapitalE{{\cal E}}
\def\ScriptN{{\mathfrak n}}
\def\ScriptE{{\mathfrak e}}
\def\ScriptP{{\mathfrak p}}

\def\RawWedge{"705E}
\def\_{"7016}
\def\overvar#1#2{\mathaccent #2 #1}

\def\cO{{\cal O}}
\def\bbD{\mathbb{D}}
\def\bbO{\mathbb{O}}
\def\bbC{\mathbb{C}}
\def\ctheta{{c_\vartheta}}
\def\stheta{{s_\vartheta}}
\def\cphi{{c_\varphi}}
\def\sphi{{s_\varphi}}
\def\v#1{{\bar{v}_{#1}}}
\def\n#1{{\hat{n}_{#1}}}
\def\phatd#1{{\hat{p}_{#1}}}
\def\p#1{{p^{\bar #1}}}
\def\a{\alpha}
\def\b{\beta}
\def\c{\gamma}
\def\d{\delta}
\def\e{\epsilon}

\begin{abstract}
Many astrophysical phenomena exhibit relativistic radiative flows. While velocities in excess of $v \sim 0.1c$ can occur in these systems, 
it has been common practice to approximate radiative transfer to $\cO(v/c)$. 
In the case of neutrino transport in core-collapse supernovae, this approximation gives rise to an inconsistency between the lepton number transfer and lab frame energy transfer, which
 have different $\cO(v/c)$ limits. 
A solution used in spherically symmetric $\cO(v/c)$ simulations has been to retain, for energy accounting purposes, the $\cO(v^2/c^2)$ terms in the lab frame energy transfer equation that arise from the $\cO(v/c)$ neutrino number transport equation. Avoiding the proliferation of such ``extra'' $\cO(v^2/c^2)$ terms in the absence of spherical symmetry
motivates a special relativistic formalism, which we exhibit in coordinates sufficiently general to encompass Cartesian, spherical, and cylindrical coordinate systems.
\end{abstract}

\pacs{97.60.Bw, 95.30.Jx, 47.70.-n}

\maketitle

\section{Motivation}

The requirement of energy- and angle-dependent neutrino radiative transfer makes a core-collapse supernova a seven-dimensional problem: In addition to the four dimensions of spacetime, the neutrino distribution functions depend on a three-dimensional momentum space. 
Energy- and angle-dependent neutrino transport has been employed in spherically symmetric simulations by a number of groups \cite{rampp00,mezzacappa01,thompson03}. In the spatially multidimensional case, energy- and angle-dependent transport has been implemented in an approximate `ray-by-ray' fashion that, except for advection, ignores lateral transport  \cite{buras03}. Efforts aimed at unapproximated versions are underway as well \cite{livne04,cardall04}. For more information on the need for detailed neutrino transport, and 
the history of approximate treatments and resulting insights into the explosion mechanism, see for example the reviews in Refs. \cite{mezzacappa04,cardall05}.

With one exception in spherical symmetry \cite{liebendoerfer01a}, all of this work has used equations of neutrino radiative transfer approximated to $\cO(v/c)$; but this gives rise to an inconsistency associated with energy conservation, a crucial check of a simulation's accuracy that has implications for claims about the explosion mechanism. The difficulty is that energy conservation can only be assessed in the ``lab frame'' (that is, by an inertial observer, necessarily at rest at infinity in the general relativistic case)
\cite{liebendoerfer01b,liebendoerfer04}, but the $\cO(v/c)$ limit of the lab-frame energy transfer equation 
is not equivalent to the $\cO(v/c)$ limit of the energy transfer equation in a reference frame comoving with the background fluid \cite{mezzacappa04}. To see this, note first that lepton number conservation is based on the Boltzmann equation, which expresses the transport of neutrino number:
\begin{equation}
\mathbb{L}[f]=\mathbb{C}[f], \label{number}
\end{equation}
where $\mathbb{L}[f]$ is the Liouville operator and $\mathbb{C}[f]$ is the collision integral \cite{lindquist66,ehlers71,mezzacappa89,cardall03}. A commonly used energy transfer equation is Eq.~(\ref{number}) multiplied by $\e$, the neutrino energy measured in the comoving frame:
\begin{equation}
\e\,\mathbb{L}[f]=\e\,\mathbb{C}[f]. \label{comovingEnergy}
\end{equation}
However, fully conservative forms are obtained in flat spacetime \cite{cardall03} (and even in general relativity, in the case of spherical symmetry \cite{liebendoerfer01b,liebendoerfer04}) by multiplying Eq.~(\ref{number}) by the particle energy measured by a lab frame
observer. 
To $\cO(v/c)$, this lab frame energy is $\e(1+\n{i}\v{i})$, where $\e\,\n{i}$ are the components of particle momentum in the comoving frame, and $\v{i}$ are the fluid velocity components measured in the lab frame. Hence the lab frame energy transfer equation is
\begin{equation}
\e(1+\n{i}\v{i})\,\mathbb{L}[f]=\e(1+\n{i}\v{i})\,\mathbb{C}[f]. \label{labEnergy}
\end{equation}
When $f$ is taken to depend on 
neutrino energy and angles measured in the comoving frame, $\mathbb{L}[f]$ contains velocity-dependent terms representing Doppler shifts and angular aberrations.
 But the $\cO(v/c)$ limit of $\mathbb{L}[f]$ gives rise to terms of $\cO(v^2/c^2)$ on the left-hand side of Eq.~(\ref{labEnergy}); dropping these yields an $\cO(v/c)$ lab frame energy transfer equation that is not equivalent to the $\cO(v/c)$ limit of Eqs.~(\ref{number}) and (\ref{comovingEnergy}). 

There have been two basic approaches to the problem of computing both energy and lepton number transfer with energy- and angle-dependent neutrino transport, but only one group has taken special care with global energy conservation. Two groups \cite{burrows00,rampp02} have employed a variable Eddington factor approach (which might also be called an ``iterated moment method'' \cite{cardall04}) to solve Eq.~(\ref{comovingEnergy}) for the neutrino energy distributions in the comoving frame. The integral of Eq.~(\ref{comovingEnergy}) over momentum space provides a heat source to the fluid internal energy, and the lepton number transfer follows easily from the momentum space integral of Eq.~(\ref{number}). However, consistency with Eq.~(\ref{labEnergy})---which bears on global energy conservation---has not been addressed by these groups. In the other approach \cite{mezzacappa93b,mezzacappa99,liebendoerfer04,mezzacappa04} (which might be called a ``direct method'' \cite{cardall04}), a complete discretization of Eq.~(\ref{number}) is solved directly. Its integral over momentum space yields the lepton number transferred to the fluid, and the source for fluid internal energy follows from the momentum space integral of Eq.~(\ref{comovingEnergy}). However, in this case special care has been taken to ensure that the discretization of Eq.~(\ref{number}) is consistent with Eq.~(\ref{labEnergy}), so that global energy conservation is maintained at a level capable of resolving with confidence the phenomenon of greatest interest: an explosion with kinetic energy $\sim 10^{51}$ erg.

This explosion energy of $\sim 10^{51}$ erg must be discriminated against a background of $\sim 10^{53}$ erg, the natural energy scale of the system. This larger reservoir of gravitational potential energy is released during collapse and transmuted  into neutrinos, preceding the fractional energy return to the fluid in the form of neutrino heating that leads eventually to kinetic expansion. This transmutation is a genuine radiation hydrodynamics problem, in which the neutrinos and matter are tightly coupled; hence it is difficult to justify an argument that errors of $\sim 10^{51}$ erg in the global neutrino energy are irrelevant to the magnitude of errors in the fluid kinetic energy. Maintenance of excursions in global energy to an order of magnitude below $\sim 10^{51}$ erg corresponds to a precision of one part in $10^{3}$. A typical simulation will be carried out over $\sim 10^{5-6}$ time steps, which requires that energy be conserved systematically to better than one part in $10^{8-9}$ per time step. This is a severe requirement, one that is very difficult to satisfy in a realistic supernova model. 


Such precision is a necessary (but not sufficient) condition for confidence that simulation outcomes represent reality. Even assuming all the relevant physics has been included, it is true that conservation of lepton number and energy are no guarantee that a model is correct: It is possible that a model could conserve total energy, while not accurately partitioning the total energy among kinetic, internal, gravitational, and other components. Nevertheless, a model that does {\it not} conserve lepton number and energy fails an obvious measure of quality control. How should we interpret the prediction of a $\sim 10^{51}$ erg explosion in a model where the total energy varies during the course of the simulation by $\sim 10^{51}$ erg or more?


It is true that in spherical symmetry all three groups agree qualitatively on a failure to explode \cite{rampp00,mezzacappa01,thompson03}, and that two of the groups have been shown to compare favorably in quantitative terms during some simulation epochs \cite{liebendoerfer05}; but the need for a careful treatment of neutrino transport that conserves global energy remains necessary as the field moves towards two and three space dimensions. For one thing, assessing the mutual validity of the simulations required a comparison with the 
transport treatment 
that underwent continued improvement until the monitored global energy was conserved at the required level \cite{liebendoerfer04}. Moreover, the simulation comparisons showed some nontrivial differences during the outward trajectory and initial stall of the shock. How these differences might play out in the multidimensional case---which is closer to exploding---remains to be seen. The history of qualitative disagreement in the outcome of simulations by different groups in multiple space dimensions \cite{burrows95,mezzacappa98b,fryer02,buras03}---not to mention the fact that experience has demonstrated that small variations in neutrino transport treatments can sometimes lead to qualitative changes in outcome \cite{janka03}---indicates that care and caution are called for (see also the review in Ref. \cite{cardall05}).


Because of the distinct $\cO(v/c)$ limits in energy and lepton number transfer, a treatment valid to all orders in $v/c$ is arguably simpler. Infall velocities of $0.2c-0.4c$ are sufficiently high during collapse  that the $\cO(v^2/c^2)$ terms in Eq.~(\ref{labEnergy}) are not negligible, and must be retained for the purposes of global energy accounting \cite{liebendoerfer04,mezzacappa04}. This indicates, of course, that the $\cO(v/c)$
limit is quantitatively inadequate, but there is also a practical reason to go further: In the relativistic case, Eq.~(\ref{labEnergy}) actually simplifies, with the ``extra'' $\cO(v^2/c^2)$ terms disappearing (see Refs. \cite{liebendoerfer01b,liebendoerfer04} for the spherically symmetric case, and Ref. \cite{cardall03} and the present work for the spatially multidimensional case). In two and three space dimensions, the number of ``extra'' $\cO(v^2/c^2)$ terms is greatly multiplied, rendering this practical dividend of a relativistic formalism even more attractive. Due to the absence of gravitational radiation in spherical symmetry, general relativistic simulations \cite{liebendoerfer01a,liebendoerfer04} are not qualitatively more difficult than those done in the Newtonian+$\cO(v/c)$ limit. While supernova simulations are not yet of sufficient maturity to take on the challenge of full numerical general relativity in two and three space dimensions---and will thus have a quantitative deficiency at some level---a special relativistic formalism nevertheless avoids the ``extra'' $\cO(v^2/c^2)$ terms. Enhancements of Newtonian gravity can then capture some of the general relativistic effects \cite{cardall01,rampp02,marek05}.

This paper is organized as follows. In Sec. \ref{sec:coordinates}, we present the general relativistic Boltzmann transport equation and its conservative variants, and describe our space and momentum space coordinate choices. Expressions in the transport equations are specialized to flat spacetime in Sec. \ref{sec:specialization}, in coordinates sufficiently general to encompass Cartesian, spherical, and cylindrical coordinate systems.
Section \ref{sec:summary} summarizes our results, and mentions a couple of other astrophysical examples requiring relativistic radiation transport. An appendix details the relationship between the number- and lab-frame energy-conservative equations---a relationship that ought to inform the construction of finite-difference representations. 

\section{Preparation}
\label{sec:coordinates}

We begin by displaying three compact forms of the transport equation---valid even in general relativity---that involve global spacetime coordinates and momentum variables reckoned in a frame comoving with the fluid with which the transported particles interact. The Boltzmann equation for electrically neutral particles is \cite{lindquist66,ehlers71,mezzacappa89,cardall03}
\begin{equation}
p^{\mu}{\partial f\over\partial x^\mu} -
{\Gamma^{\hat\jmath}}_{\hat\mu\hat\nu}p^{\hat\mu}p^{\hat\nu}
{\partial u^{\tilde\imath}\over\partial p^{\hat\jmath}}{\partial f\over\partial 
u^{\tilde\imath}} = \mathbb{C}[f]. \label{boltzmann}
\end{equation}
Equivalent in content are the conservative expressions for particle number and energy-momentum conservation \cite{cardall03}:
\begin{widetext}
\begin{equation}
{1\over\sqrt{-g}}{\partial\over\partial x^\mu}\left(\sqrt{-g}\,
p^{\mu}f\right) - 
\epsilon
\left|\det\left(\partial{\bf p}\over \partial{\bf u}\right)\right|^{-1} 
{\partial\over\partial u^{\tilde\imath}}\left({1\over\epsilon}
\left|\det\left(\partial{\bf p}\over \partial{\bf u}\right)\right|
{\partial u^{\tilde\imath}\over\partial p^{\hat\jmath}}\,
{\Gamma^{\hat\jmath}}_{\hat\mu\hat\nu}\,{p^{\hat\mu}p^{\hat\nu}}
f\right) = \mathbb{C}[f],
\label{numberConservative}
\end{equation}
\begin{equation}
{1\over\sqrt{-g}}{\partial\over\partial x^\mu}\left(\sqrt{-g}\,
p^{\rho}p^{\mu}f\right) - 
\epsilon
\left|\det\left(\partial{\bf p}\over \partial{\bf u}\right)\right|^{-1} 
{\partial\over\partial u^{\tilde\imath}}\left({1\over\epsilon}
\left|\det\left(\partial{\bf p}\over \partial{\bf u}\right)\right|
{\partial u^{\tilde\imath}\over\partial p^{\hat\jmath}}\,
{\Gamma^{\hat\jmath}}_{\hat\mu\hat\nu}\,p^{\hat\mu}p^{\hat\nu}{p^{\rho}}
 f\right) = p^{\rho}\mathbb{C}[f]-{\Gamma^\rho}_{\nu\mu}p^\nu p^\mu,
\label{energyConservative}
\end{equation}
\end{widetext}
which consist of divergences in spacetime and momentum space. 

The quantity at the heart of Eqs.~(\ref{boltzmann})-(\ref{energyConservative}) is the scalar distribution $f$, a function of
spacetime coordinates $x^\mu$ and three-momentum coordinates
$u^{\tilde\imath}$. (Greek and latin letters are spacetime and space indices
respectively. Unadorned indices indicate components in a global coordinate
basis, barred indices ($\bar{\ }$) denote an orthonormal ``lab frame'' basis, hatted indices $(\hat{\ })$ indicate an
orthonormal basis
comoving with the fluid with which the particles interact (``comoving frame''), and
indices with a tilde $(\tilde{\ })$ denote a comoving frame momentum space coordinate basis.) 
The scalar distribution function $f$ gives the number of particles $dN$
in an invariant spacetime 3-volume element $dV$ and invariant
momentum space volume element $dP$ \cite{lindquist66}:
\begin{equation}
dN = f(x,{\bf p})\,(-w\cdot p)\,dV\,dP. \label{scalarDistribution}
\end{equation}
The quantities $x$, $w$, and $p$ are 4-vectors, and ${\bf p}$ is
the spatial 3-vector portion of $p$. The unit 4-vector $w$ is timelike, and defines the
orientation of $dV$:
\begin{equation}
dV = \sqrt{-g}\,\epsilon_{\mu\nu\rho\sigma}w^\mu\,d_1x^\nu\,d_2x^\rho\,
  d_3x^\sigma,
\end{equation}
where $g$ is the determinant of the metric tensor (taken to have
signature $+++-$) and $\epsilon_{\mu\nu\rho\sigma}$ is the Levi-Civita
alternating symbol ($\epsilon_{0123}=+1$). The momentum space volume element is
\begin{eqnarray}
dP &=& \sqrt{-g}\,\epsilon_{ijk} {\,d_1p^i\,d_2p^j\,
    d_3p^k \over (-p_0)} \nonumber \\
  &=& {1\over \epsilon}\left|\det\left(\partial{\bf p}\over \partial{\bf u}\right)\right|
    \,du^{\tilde{1}}\,du^{\tilde{2}}\,du^{\tilde{3}}, \label{momentumElement}
\end{eqnarray}
where specialization to momentum variables $u^{\tilde\imath}$ in the comoving
frame has been made in the second line, and 
\begin{equation}
\epsilon \equiv p^{\hat 0} = \sqrt{|{\bf p}|^2 + m^2}\label{comovingMomentum0}
\end{equation}
is the particle
energy measured in the comoving frame ($m$ is the particle mass, taken to be zero for the purposes of photon or classical neutrino transport). 
The definition of $f$ in Eq.~(\ref{scalarDistribution}) makes
a convenient connection to nonrelativistic definitions of the
distribution function. For an equivalent but more geometric approach,
see Ref. \cite{ehlers71}.


We now describe in more detail the spacetime 
and momentum space coordinate systems we will use in the special relativistic elaboration of Eqs.~(\ref{boltzmann})-(\ref{energyConservative}).
In spacetime, we have a global coordinate basis. In momentum space, we invoke orthonormal ``lab frame'' and ``comoving'' spacetime bases to obtain the Cartesian spatial momentum components $p^{\hat\imath}$ measured in a comoving orthonormal frame, and then transform these to momentum space spherical coordinates $u^{\tilde\imath}$ (particle energy, and two angles specifying the momentum direction).


\begin{table}
\caption{\label{tab:coordinates}Coordinates, metric functions, and derivatives of metric functions for common coordinate systems.}
\begin{ruledtabular}
\begin{tabular}{cccccccccc}
System & $x^1$ &  $x^2$ &  $x^3$ & $a$ & $b$ & $c$ & $\displaystyle {1\over a}{\partial a\over\partial x^1}$ & $ \displaystyle {1\over b}{\partial b\over\partial x^1}$ & $\displaystyle {1\over a c}{\partial c\over\partial x^2}$\\
\hline
Cartesian & $x$ & $y$ & $z$ & 1 & 1 & 1 & 0 & 0 & 0\\
Spherical & $r$ & $\theta$ & $\phi$ & $r$ & $r$ & $\sin\theta$ & $\displaystyle {1\over r}$ & $\displaystyle {1\over r}$ & $\displaystyle {\cot\theta \over r}$\\
Cylindrical & $r$ & $z$ & $\phi$ & 1 & $r$ & 1 & 0 & $\displaystyle {1\over r}$ & 0\\ 
\end{tabular}
\end{ruledtabular}
\end{table}

We assume flat spacetime, but in order to accomodate curvilinear coordinate
systems we label our spacetime global coordinates \inlineSFinmath{$({x^{\mu }})={{\big({x^1},{x^2},{x^3},t\big)}^T}$}. (In matrix expressions of tensor components, rows and columns are ordered \inlineSFinmath${1,2,3,0}$.) The line element is
\begin{equation}
\dispSFNumberedEquationmath{{{\Mvariable{ds}}^2}={g_{\Mvariable{\mu \nu }}}{{\Mvariable{dx}}^{\InvisibleComma \mu }}{{\Mvariable{dx}}^{\nu }},}
\end{equation}
where the metric is diagonal: 
\begin{equation}
(g_{\mu\nu})={\rm diag}[1,{a^2}({x^1}),{b^2}({x^1})\,{c^2}({x^2}),-1].\label{flatMetric}
\end{equation}
The specialization of the coordinates ${x^1},{x^2},{x^3}$ and metric functions $a,b,c$ in various coordinate systems are given in Table \ref{tab:coordinates}.
Our spacetime coordinate
systems are ``lab frames,'' so that the equations we derive are Eulerian. 

The orthonormal ``lab frame'' is
accessed with the ``tetrad'' ${e^\mu}_{\bar \mu}$, which locally transform the metric into the Lorentz form $\eta_{\bar\mu \bar\nu}={\rm diag}[1,1,1,-1]$:
\begin{equation}
{e^\mu}_{\bar \mu}\,{e^\nu}_{\bar \nu}\,g_{\mu\nu}=\eta_{\bar\mu \bar\nu}.
\end{equation}
A simple choice is
\begin{equation}
\left({e^\mu}_{\bar \mu}\right)={\rm diag}[1,a^{-1}({x^1}),b^{-1}({x^1})\,c^{-1}({x^2}),1].\label{tetrad}
\end{equation}
Because this is diagonal, the inverse ${e^{\bar\mu}}_{\mu}$
simply has the diagonal entries inverted. 

The orthonormal frame comoving with the fluid---indicated by indices adorned with a hat ($\hat{\ }$)---is related to the orthonormal lab frame by a Lorentz boost, which satisfies
\begin{equation}
{\Lambda^{\bar \mu}}_{\hat \mu }{\Lambda^{\bar \nu}}_{\hat \nu }\, \eta_{\bar\mu \bar\nu} = \eta_{\hat\mu \hat\nu},
\end{equation}
where
\begin{eqnarray}
{\Lambda^{\bar\imath}}_{\hat\jmath }&=& \d_{ij}+{\c^2\over \c+1}\v{i}\v{j}, \label{boost1}\\
{\Lambda^{\bar\imath}}_{\hat 0 }&=& \c\v{i}, \\
{\Lambda^{\bar 0}}_{\hat\imath}&=& \c\v{i}, \label{boost3}\\
{\Lambda^{\bar 0}}_{\hat 0 }&=& \c, \label{boost0}
\end{eqnarray}
and 
\begin{equation}
\c\equiv\left[1-\left(\v{1}\right)^2-\left(\v{2}\right)^2-\left(\v{3}\right)^2\right]^{-1/2}. \label{gamma}
\end{equation}
The inverse ${\Lambda^{\hat \mu}}_{\bar \mu }$---the boost from the lab frame to the comoving frame---is obtained by taking $\v{i}\rightarrow -\v{i}$.  
The bars on the velocities are a reminder that these are the 3-velocity components in the orthonormal lab frame coordinate system (they are not spatial components of a 4-vector). 

We also define a composite
transformation (technically, also a tetrad) from the orthonormal comoving frame to the global coordinate frame: 
\begin{equation}
{{\cal L}^{\mu}}_{\hat\mu}{{\cal L}^{\nu}}_{\hat\nu}\,g_{\mu\nu}=\eta_{\hat\mu \hat\nu},
\end{equation}
where
\begin{equation}
{{\cal L}^{\mu}}_{\hat\mu}={e^\mu}_{\bar \mu}{\Lambda^{\bar \mu}}_{\hat \mu }.\label{combinedTransformation}
\end{equation}
The inverse ${{\cal L}^{\hat\mu}}_{\mu}={\Lambda^{\hat \mu}}_{\bar \mu }{e^{\bar\mu}}_{ \mu}$ is the transformation from the comoving frame to the orthonormal lab frame.

Finally, the neutrino 4-momentum is described in terms of its comoving frame components, \inlineSFinmath${\Big({p^{\overvar{\mu }{\RawWedge }}}\Big)=\Big({p^{\overvar{1}{\RawWedge
}}},{p^{\overvar{2}{\RawWedge }}},{p^{\overvar{3}{\RawWedge }}},{p^{\overvar{0}{\RawWedge }}}\Big)^T}$. Only three momentum variables are independent;
we choose polar coordinates in momentum space, which for massless particles may be expressed
\begin{eqnarray}
{p^{\overvar{1}{\RawWedge }}}&=&\epsilon\, \multsp \Mvariable{\cos}\vartheta  \equiv \e \, \n{1},
\label{comovingMomentum1}\\
{p^{\overvar{2}{\RawWedge }}}&=&\epsilon\, \multsp \Mvariable{\sin}\vartheta\, \Mvariable{\cos}\varphi  \equiv \e \, \n{2},
\\
{p^{\overvar{3}{\RawWedge }}}&=&\epsilon\, \multsp \Mvariable{\sin}\vartheta\, \Mvariable{\sin}\varphi  \equiv \e \, \n{3}, 
\label{comovingMomentum3}
\end{eqnarray}
where the hat on $\n{i}$ 
indicates that this 3-vector 
pertains to the orthonomal comoving frame. 
(Note that $\n{i}$ is not spatial portion of a 4-vector. With both $\n{i}$ and $\v{i}$ we employ a repeated index summation convention that does not require one index up and one index down for these non-spacetime-tensor quantities; their indices are always down.)

In summary, the neutrino radiation field is a function of the variables 
\inlineSFinmath{$({x^{\mu }})={{\big({x^1},{x^2},{x^3},t\big)}^T}$}
and \inlineSFinmath{$({u^{\tilde\imath}})={{\big({\e},{\vartheta},{\varphi}\big)}^T}$}.

\section{Elaboration}
\label{sec:specialization}

With these specific coordinate choices, we are ready to specialize the expressions appearing in Eqs.~(\ref{boltzmann}-\ref{energyConservative}). Three expressions are relatively simple, and one is complicated.

The simplest are the two determinant expressions in Eqs.~(\ref{numberConservative}) and (\ref{energyConservative}). The determinant of the metric tensor is $g$, so from Eq.~(\ref{flatMetric}),
\begin{equation}
\sqrt{-g}=a b c.
\end{equation}
The Jacobian determinant of the momentum space coordinate change is
\begin{equation}
\left|\det\left(\partial{\bf p}\over \partial{\bf u}\right)\right|=\e^2\sin\vartheta,
\end{equation}
which follows from Eqs.~(\ref{comovingMomentum1}-\ref{comovingMomentum3}).

Also simple 
are the global coordinate basis momentum components $p^\mu={{\cal L}^{\mu}}_{\hat\mu}p^{\hat\mu}$
appearing in the 
transport equations (the composite transformation ${{\cal L}^{\mu}}_{\hat\mu}$ is given by Eq.~(\ref{combinedTransformation})). From the boost, Eqs.~(\ref{boost1}-\ref{boost0}), and comoving frame momentum components in Eqs.~(\ref{comovingMomentum0}), (\ref{comovingMomentum1}-\ref{comovingMomentum3}), the orthonormal lab frame momentum components $\p{\mu}={\Lambda^{\bar \mu}}_{\hat \mu }\,p^{\hat\mu}$ are
\begin{eqnarray}
p^{\bar\imath}&=&\e\left[ \n{i}+	\c\v{i}\left(1+{\c\over\c+1}\,\n{j}\v{j}\right)\right], \label{labMomentumi}\\
\p{0}&=&\c\,\e\left(1+\n{i}\v{i}\right). \label{labMomentum0}
\end{eqnarray} 
The coordinate basis momentum components $p^{\mu}={e^{\mu}}_{\bar \mu }p^{\bar\mu}$ are
\begin{eqnarray}
p^{1}&=&{e^{1}}_{\bar\mu}\,p^{\bar\mu}=\p{1}, \label{coordMomentum1}\\
p^{2}&=&{e^{2}}_{\bar\mu}\,p^{\bar\mu}={1\over a}\,\p{2}, \\
p^{3}&=&{e^{3}}_{\bar\mu}\,p^{\bar\mu}={1\over b c}\,\p{3}, \\
p^0 &=& {e^{0}}_{\bar\mu}\,p^{\bar\mu}=\p{0},\label{coordMomentum0}
\end{eqnarray}
where we have used
Eq.~(\ref{tetrad}).

The final expression to specialize from Eqs.~(\ref{boltzmann}-\ref{energyConservative}) is $({\partial u^{\tilde\imath}/\partial p^{\hat\jmath}})\,
{\Gamma^{\hat\jmath}}_{\hat\mu\hat\nu}\,{p^{\hat\mu}p^{\hat\nu}}$, which, being complicated, constitutes the computational burden of this paper. The first factor can be obtained by computing the Jacobian matrix $(\partial p^{\hat\imath}/\partial u^{\tilde j})$ from Eqs.~(\ref{comovingMomentum1})-(\ref{comovingMomentum3}) and taking the matrix inverse; the result is
\begin{equation}
\left({\partial u^{\tilde\imath}\over\partial p^{\hat\jmath}}\right) =\pmatrix{ 
\cos\vartheta& \sin\vartheta\cos\varphi & \sin\vartheta\sin\varphi \cr 
\displaystyle -{\sin\vartheta\over\e} &\displaystyle {\cos\vartheta\cos\varphi\over\e} & \displaystyle {\cos\vartheta\sin\varphi\over\e} \cr 0 &\displaystyle -{\sin\varphi\over\e\sin\vartheta} & \displaystyle {\cos\varphi\over\e\sin\vartheta}}. \label{dudp}
\end{equation}
The connection coefficients
in the orthonormal comoving basis are
\begin{equation}
{\Gamma^{\hat\mu}}_{\hat\nu\hat\rho} = {\mathcal{L}^{\hat\mu}}_{\mu}{\mathcal{L}^\nu}_{\hat\nu}{\mathcal{L}^\rho}_{\hat\rho}\, {\Gamma^{\mu}}_{\nu\rho} +
{\mathcal{L}^{\hat\mu}}_{\mu}{\mathcal{L}^\rho}_{\hat\rho}
{\partial {\mathcal{L}^\mu}_{\hat\nu}\over\partial x^\rho},
\label{connectionComoving}
\end{equation} 
where
\begin{equation}
{\Gamma^\mu}_{\nu\rho} = {1\over 2}g^{\mu\sigma}\left({\partial g_{\sigma\nu}
\over\partial x^\rho} + {\partial g_{\sigma\rho}\over\partial x^\nu} - 
{\partial g_{\nu\rho}\over\partial x^\sigma}\right)
\end{equation}
are the coordinate basis connection coefficients, and the ${\cal L}$ transformations are given by Eq.~(\ref{tetrad}) and subsequent text, Eqs.~(\ref{boost1}-\ref{gamma}) and subsequent text, and Eq.~(\ref{combinedTransformation}) and subsequent text. 
We note the following combination:
\begin{eqnarray}
{\Gamma^{\hat\jmath}}_{\hat\mu\hat\nu}\,{p^{\hat\mu}p^{\hat\nu}} &=&
\left({\mathcal{L}^{\hat\jmath}}_{\mu}{\Gamma^\mu}_{\nu\rho}+
{\mathcal{L}^{\hat\jmath}}_{\mu}{\partial{\mathcal{L}^{\mu}}_{\hat\nu}\over\partial x^\rho}{\mathcal{L}^{\hat \nu}}_{\nu}\right)p^\nu p^\rho \nonumber \\
&=&\left({\Gamma^{\hat\jmath}}_{\hat\mu\hat\nu}\,{p^{\hat\mu}p^{\hat\nu}}\right)_{abc} + \left({\Gamma^{\hat\jmath}}_{\hat\mu\hat\nu}\,{p^{\hat\mu}p^{\hat\nu}}\right)_{\v{1}\v{2}\v{3}}.\label{combination}
\end{eqnarray}
The first term involves derivatives of the metric functions $a$, $b$, and $c$:
\begin{eqnarray}
\left({\Gamma^{\hat\jmath}}_{\hat\mu\hat\nu}\,{p^{\hat\mu}p^{\hat\nu}}\right)_{abc}&=&
{\mathcal{L}^{\hat\jmath}}_{\mu}{\Gamma^\mu}_{\nu\rho}p^\nu p^\rho +
{\Lambda^{\hat\jmath}}_{\bar\mu}{e^{\bar\mu}}_\mu{\partial{e^{\mu}}_{\bar\nu}\over\partial x^\rho}p^{\bar\nu} p^\rho \nonumber \\
&=& A^{\hat\jmath}+B^{\hat\jmath}+C^{\hat\jmath},\label{abc}
\end{eqnarray}
where
\begin{eqnarray}
A^{\hat\jmath} &=& \left({\Lambda^{\hat\jmath}}_{\bar 2}\p{1} - {\Lambda^{\hat\jmath}}_{\bar 1}\p{2}\right){\p{2}\over a}{\partial a\over\partial x^1}, \label{afactor}\\
B^{\hat\jmath} &=& \left({\Lambda^{\hat\jmath}}_{\bar 3}\p{1} - {\Lambda^{\hat\jmath}}_{\bar 1}\p{3}\right){\p{3}\over b}{\partial b\over\partial x^1}, \label{bfactor}\\
C^{\hat\jmath} &=& \left({\Lambda^{\hat\jmath}}_{\bar 3}\p{2} - {\Lambda^{\hat\jmath}}_{\bar 2}\p{3}\right){\p{3}\over a c}{\partial c\over\partial x^2}.\label{cfactor}
\end{eqnarray}
See Table \ref{tab:coordinates} for explicit values of the metric derivatives in various coordinate systems.
The second term of Eq.~(\ref{combination}) involves derivatives of the velocity components:
\begin{eqnarray}
\left({\Gamma^{\hat\jmath}}_{\hat\mu\hat\nu}\,{p^{\hat\mu}p^{\hat\nu}}\right)_{\v{1}\v{2}\v{3}}&=&
{\Lambda^{\hat\jmath}}_{\bar\mu}{\partial{\Lambda^{\bar\mu}}_{\hat\nu}\over\partial x^\mu}p^{\hat\nu} p^\mu \nonumber \\
&=& V^{\hat\jmath},
\end{eqnarray}
where 
\begin{eqnarray}
V^{\hat\jmath} &=& \e\left[\c{\partial\v{j}\over\partial x^\mu}+{\v{j}\over \c+1}{\partial\c\over\partial x^\mu}+\right.\nonumber\\
& &\left.{\c^2 \over \c+1}\left(\v{k}{\partial\v{j}\over\partial x^\mu}-\v{j}{\partial\v{k}\over\partial x^\mu} \right)\n{k}\right]{e^{\mu}}_{\bar\mu}\,p^{\bar\mu}.\label{vfactor}
\end{eqnarray}
We note that this contribution to Eq.~(\ref{combination}) will obtain even in general relativity (with suitable expressions for the tetrad ${e^{\mu}}_{\bar\mu}$).

Particularly for use in approaches employing angular moments, it may be worth noting that the ${\tilde\imath}=\e$ elements of the multiplication by $\partial u^{\tilde\imath}/\partial p^{\hat\jmath}$ simplify noticeably:
\begin{eqnarray}
{\partial u^{\e}\over\partial p^{\hat\jmath}}\; A^{\hat\jmath}&=& \c\,\e\left(\n{2}\v{1}-\n{1}\v{2}\right){\p{2}\over a}{\partial a\over\partial x^1}, \label{moment1}\\
{\partial u^{\e}\over\partial p^{\hat\jmath}}\; B^{\hat\jmath}&=& \c\,\e\left(\n{3}\v{1}-\n{1}\v{3}\right){\p{3}\over b}{\partial b\over\partial x^1}, \\
{\partial u^{\e}\over\partial p^{\hat\jmath}}\; C^{\hat\jmath}&=& \c\,\e\left(\n{3}\v{2}-\n{2}\v{3}\right){\p{3}\over a c}{\partial c\over\partial x^2}. \\
{\partial u^{\e}\over\partial p^{\hat\jmath}}\; V^{\hat\jmath}&=& 
\e\,\n{j}\left(\c{\partial\v{j}\over\partial x^\mu}+{\v{j}\over \c+1}{\partial\c\over\partial x^\mu}\right){e^{\mu}}_{\bar\mu}\,p^{\bar\mu} \label{moment4}.
\end{eqnarray}
Beyond this, more explicit expressions of the $\tilde\imath=\vartheta,\varphi$ elements are not particularly illuminating; and even in computer codes it may be worthwhile, for a reason mentioned in the appendix, to use the somewhat more general expressions in the previous paragraph.

\section{Summation}
\label{sec:summary}

To summarize, the nonconservative (Eq.~(\ref{boltzmann})), number-conservative (Eq.~(\ref{numberConservative})), and energy-conservative ($\rho=0$ component of Eq.~(\ref{energyConservative})) special relativistic transport equations for the massless particle scalar distribution function $f$, a function of global spacetime coordinates \inlineSFinmath{$({x^{\mu }})={{\big({x^1},{x^2},{x^3},t\big)}^T}$}
and comoving-frame energy and angles \inlineSFinmath{$({u^{\tilde\imath}})={{\big({\e},{\vartheta},{\varphi}\big)}^T}$}---with metric functions $a$, $b$, and $c$ sufficiently general to allow for Cartesian, spherical, and cylindrical coordinates---are
\begin{widetext}
\begin{equation}
{e^{\mu}}_{\bar\mu}\,p^{\bar\mu}{\partial f\over\partial x^\mu} -
\left(A^{\hat\jmath} + 
	B^{\hat\jmath} +
	C^{\hat\jmath} +
	V^{\hat\jmath}\right)
{\partial u^{\tilde\imath}\over\partial p^{\hat\jmath}}
{\partial f\over\partial 
u^{\tilde\imath}} = \mathbb{C}[f], \label{boltzmann2}
\end{equation}
\begin{equation}
{1\over a b c}{\partial\over\partial x^\mu}\left(a b c\,{e^{\mu}}_{\bar\mu}\,p^{\bar\mu}f\right) - 
{1\over\epsilon\sin\vartheta}
{\partial\over\partial u^{\tilde\imath}}\left[\epsilon\sin\vartheta\,
{\partial u^{\tilde\imath}\over\partial p^{\hat\jmath}}
\left(A^{\hat\jmath} + 
	B^{\hat\jmath} +
	C^{\hat\jmath} +
	V^{\hat\jmath}\right)
f\right] = \mathbb{C}[f],
\label{numberConservative2}
\end{equation} 
\begin{equation}
{1\over a b c}{\partial\over\partial x^\mu}\left(a b c\,{e^{\mu}}_{\bar\mu}\,p^{\bar\mu}\,p^{\bar 0}f\right) - 
{1\over\epsilon\sin\vartheta}
{\partial\over\partial u^{\tilde\imath}}\left[\epsilon\sin\vartheta\,
{\partial u^{\tilde\imath}\over\partial p^{\hat\jmath}}
\left(A^{\hat\jmath} + 
	B^{\hat\jmath} +
	C^{\hat\jmath} +
	V^{\hat\jmath}\right)p^{\bar 0}
f\right] = p^{\bar 0}\, \mathbb{C}[f],
\label{energyConservative2}
\end{equation} 
\end{widetext}
where $A^{\hat\jmath}$, $B^{\hat\jmath}$, $C^{\hat\jmath}$, and $V^{\hat\jmath}$ are given by Eqs.~(\ref{afactor}-\ref{cfactor}), (\ref{vfactor}). We remind that in all these expressions, care must be taken to distinguish between the momentum components in different frames, given by
Eqs.~(\ref{comovingMomentum0}), (\ref{comovingMomentum1}-\ref{comovingMomentum3}), and (\ref{labMomentumi}-\ref{coordMomentum0}).
The expression ${\partial u^{\tilde\imath}/\partial p^{\hat\jmath}}$ is given by Eq.~(\ref{dudp}).
Finally, $\c$ (given by Eq.~(\ref{gamma}))
is the Lorentz factor of the boost (Eqs.~(\ref{boost1}-\ref{boost0})) between the orthonormal ``lab'' and ``comoving'' frames. The expressions in Eq.~(\ref{moment1}-\ref{moment4}) will be convenient for use in angular moment formalisms (such as variable Eddington factor approaches).

Our final equations are valid in flat spacetime. In Eq.~(\ref{energyConservative2}), we note that ${\Gamma^0}_{\mu\nu}$ in Eq.~(\ref{energyConservative}) has vanished in flat spacetime, but (in general) would not vanish in curved spacetime. Also, $p^{\bar 0}$ would have to be replaced by the more general expression ${e^{0}}_{\bar\mu}\,p^{\bar\mu}$. But even in curved spacetime, the $V^{\hat \jmath}$ term will be the same, except for a different tetrad ${e^{\mu}}_{\bar\mu}$ in Eq.~(\ref{vfactor}). Finally, the major difference in curved spacetime will be the replacement of Eq.~(\ref{abc}) with terms arising from the more general metric.

In the appendix we examine in detail the relationship between these number and lab frame energy transport equations, highlighting expressions that should cancel in a finite-differenced representation.
Our finite-differenced representation of the special relativistic neutrino number transport equation and the results of core-collapse simulations using it will be reported separately,
but for now we can mention some salient points regarding our approach to finite differencing. First, because we must be able to accurately reproduce neutrino diffusion inside the nascent neutron star while retaining stable free streaming outside, our spatial differencing interpolates between centered differencing (good for diffusion) and ``upwind'' differencing (stable free streaming), according to the neutrino mean free path \cite{mezzacappa93b,liebendoerfer04}. Second, also important for accuracy in the diffusion limit, certain terms cancel between the space and momentum space divergences in the homogeneous and isotropic limit; differencings of $A^{\hat\jmath}$, $B^{\hat\jmath}$, and $C^{\hat\jmath}$ are chosen that respect this \cite{mezzacappa93b,liebendoerfer04}. Third, the conservative formulation---with its divergences in space and momentum space---invites a conservative differencing, patterned after the mathematical definition of the divergence: a sum over cell faces, divided by the cell volume. Global conservation then follows naturally, following the discrete version of the divergence theorem. A discrete volume integral is obtained by multiplying the divergence in each cell by the cell volume; the contributions on faces shared by adjacent cells then cancel exactly. Fourth, because neutrino transport in supernovae requires both number and energy conservation, we discretize the number-conservative Eq. (\ref{numberConservative2}), and work out algebraically the differencings required to ensure consistency with lab-frame energy conservation as encoded in Eq. (\ref{energyConservative2}). This involves, for example, the differencing of velocity gradients in $V^{\hat\jmath}$ being tied to the differencing chosen for the space divergence \cite{liebendoerfer04,mezzacappa04}.

While our parochial concern is neutrino transport in supernovae, relativistic photon transport is important in many astrophysical systems. Compton scattering is analyzed in detail in Refs. \cite{psaltis97,psaltis01}, where it is concluded that ${\cal O}(v/c)$ transfer equations are inadequate even for electron bulk velocities $v \sim 0.1c$. This work was applied to spherically symmetric accretion \cite{psaltis01}. Obvious examples in which multiple space dimensions are important are gamma-ray bursts and ``superluminal'' galactic jets. A ray-tracing approach (as opposed to the grid-based discretization we intend to employ) in a background stationary Kerr metric showed that both kinematic and spacetime curvature effects are important in photon transfer in accretion disks and tori surrounding black holes \cite{fuerst04}. Another example is photon transport in supernovae, necessary for the interpretation of spectra and lightcurves.
It has been shown in spherically symmetric simulations that all special
relativistic terms must be included to accurately model the spectra of the
modestly relativistic ($v \sim 0.1c$) ejecta of supernovae \cite{baron96}.
This fully relativistic formulation has been routinely used to evaluate both
spherically symmetric explosion models and observed objects of all supernova types
\cite{nugent97,baron99,lentz01,mitchell02}. Observational indications that exceptional supernovae may produce ejecta that is both relativistic and  jet-like \cite{ghirlanda04} suggest that similarly
relativistic completeness in multidimensional modeling of these objects' spectra and lightcurves will be necessary.

\appendix*
\section{}

It is apparent from the right-hand sides of Eqs.~(\ref{numberConservative2}) and (\ref{energyConservative2}) that the latter is simply $p^{\bar 0}$ times the former, but it is not obvious how this works out on the left-hand side. Because $p^{\bar 0}$ is inside the spacetime and momentum space divergences in Eq.~(\ref{energyConservative2}), the ``extra'' terms generated by pulling $p^{\bar 0}$ inside the derivatives of Eq.~(\ref{numberConservative2}) must cancel. Here we explore this cancellation in some detail.

Consider first the ``extra'' term arising from the spacetime divergence:  
\begin{equation}
{p^{\bar 0}\over a b c}{\partial\over\partial x^\mu}\left(a b c\,{e^{\mu}}_{\bar\mu}\,p^{\bar\mu}f\right) = {1\over a b c}{\partial\over\partial x^\mu}\left(a b c\,{e^{\mu}}_{\bar\mu}\,p^{\bar\mu}\,p^{\bar 0}f\right) - E_S,
\end{equation}
where
\begin{eqnarray}
E_S &=& {e^{\mu}}_{\bar\mu}\,p^{\bar\mu}f \, {\partial p^{\bar 0}\over\partial x^\mu}\\
&=& \e\,{e^{\mu}}_{\bar\mu}\,p^{\bar\mu}f\left[\c\,\n{i}{\partial\v{i}\over\partial x^\mu}+\left(1+\n{i}\v{i}\right){\partial\gamma\over\partial x^\mu}\right],\label{ES}
\end{eqnarray}
and Eq.~(\ref{labMomentum0}) has been used in the last line.

We begin consideration of the ``extra'' term resulting from the momentum divergence by noting that care must be taken in the calculation of momentum derivatives of $p^{\bar 0}$. For this purpose, uncritical use of the expression in Eq.~(\ref{labMomentum0}) is misleading. Recall that $p^{\bar 0}$ is the result of a Lorentz boost from the comoving frame; and that while $p^{\hat\mu}$ is a 4-vector, because it is on the mass shell (of a massless particle in the present case), $p^{\hat 0} = \e$ is to be considered a function of the $p^{\hat \imath}$:
\begin{eqnarray}
p^{\bar 0} &=& {\Lambda^{\bar 0}}_{\hat \mu}\, p^{\hat\mu} \\
&=& {\Lambda^{\bar 0}}_{\hat \imath}\,p^{\hat\imath}+{\Lambda^{\bar 0}}_{\hat 0}\sqrt{\left(p^{\hat 1}\right)^2+\left(p^{\hat 2}\right)^2+\left(p^{\hat 1}\right)^2}.
\end{eqnarray} 
Hence
\begin{equation}
{\partial p^{\bar 0} \over \partial u^{\tilde \imath}}={\partial p^{\bar 0} \over \partial p^{\hat \imath}}{\partial p^{\hat \imath} \over \partial u^{\tilde \imath}} = {1\over\e}\left(\e\,{\Lambda^{\bar 0}}_{\hat \imath}+{{\Lambda^{\bar 0}}_{\hat 0}\,p_{\hat\imath}}\right){\partial p^{\hat \imath} \over \partial u^{\tilde \imath}}\label{momentumDerivatives}
\end{equation}
is the needed derivative of this component.

Now we may contemplate the ``extra'' term resulting from the momentum divergence:
\begin{widetext}
\begin{equation}
-{p^{\bar 0}\over\epsilon\sin\vartheta}
{\partial\over\partial u^{\tilde\imath}}\left[\epsilon\sin\vartheta\,
{\partial u^{\tilde\imath}\over\partial p^{\hat\jmath}}
\left(A^{\hat\jmath} + 
	B^{\hat\jmath} +
	C^{\hat\jmath} +
	V^{\hat\jmath}\right)
f\right] = -{1\over\epsilon\sin\vartheta}
{\partial\over\partial u^{\tilde\imath}}\left[\epsilon\sin\vartheta\,
{\partial u^{\tilde\imath}\over\partial p^{\hat\jmath}}
\left(A^{\hat\jmath} + 
	B^{\hat\jmath} +
	C^{\hat\jmath} +
	V^{\hat\jmath}\right)p^{\bar 0}
f\right] - E_M,
\end{equation}
where
\begin{eqnarray}
E_M &=& -{\partial u^{\tilde\imath}\over\partial p^{\hat\jmath}}
\left(A^{\hat\jmath} + 
	B^{\hat\jmath} +
	C^{\hat\jmath} +
	V^{\hat\jmath}\right)f {\partial p^{\bar 0} \over \partial u^{\tilde \imath}} \\
&=&-{1\over\e}\left(\e\,{\Lambda^{\bar 0}}_{\hat \jmath}+{{\Lambda^{\bar 0}}_{\hat 0}\,p_{\hat\jmath}}\right)
\left(A^{\hat\jmath} + 
	B^{\hat\jmath} +
	C^{\hat\jmath} +
	V^{\hat\jmath}\right)f. \label{EM}
	\end{eqnarray}
\end{widetext}
In obtaining the last line, we used Eq.~(\ref{momentumDerivatives}), which resulted in a delta function as a consequence of the multiplication of the momentum space Jacobian $\partial u^{\tilde\imath}/\partial p^{\hat\jmath}$ with its inverse ${\partial p^{\hat \imath} / \partial u^{\tilde \imath}}$. In the last sentence of Sec. \ref{sec:specialization}, we referred to the possible utility of retaining---even in computer codes---our expressions involving the matrix $\partial u^{\tilde\imath}/\partial p^{\hat\jmath}$, and we are now in a position to see why: The finite-difference representation of ${\partial p^{\hat \imath} / \partial u^{\tilde \imath}}$ the right-hand side of Eq.~(\ref{momentumDerivatives}) will be determined by a choice of the finite-difference representation of the momentum space divergence; hence the finite-difference representation of $\partial u^{\tilde\imath}/\partial p^{\hat\jmath}$ can be chosen to make it precisely the matrix inverse of the (foreordained) finite-difference representation of ${\partial p^{\hat \imath} / \partial u^{\tilde \imath}}$.

We expect the cancellation $E_S + E_M =0$ in accordance with Eqs.~(\ref{numberConservative2}) and (\ref{energyConservative2}), and because $E_S$ contains no derivatives of the metric functions $a,b,c$, there must be ``internal cancellations'' in $E_M$ such that
\begin{equation}
\left(\e\,{\Lambda^{\bar 0}}_{\hat \jmath}+{{\Lambda^{\bar 0}}_{\hat 0}\,p_{\hat\jmath}}\right)
\left(A^{\hat\jmath} + 
	B^{\hat\jmath} +
	C^{\hat\jmath}\right)=0,\label{EMabc}
\end{equation}
as we now verify. From Eqs.~(\ref{afactor}-\ref{cfactor}) we see that each of the terms arising from $A^{\hat\jmath}$, $B^{\hat\jmath}$, and  $C^{\hat\jmath}$ has an overall multiplicative factor of the form
\begin{equation}
\left(\e\,{\Lambda^{\bar 0}}_{\hat \jmath}+{{\Lambda^{\bar 0}}_{\hat 0}\,p_{\hat\jmath}}\right)
\left({\Lambda^{\hat\jmath}}_{\bar \imath}\,\p{k} - {\Lambda^{\hat\jmath}}_{\bar k}\,\p{\imath}\right). \label{EMabc2}
\end{equation}
We note that (for example) 
\begin{eqnarray}
{\Lambda^{\bar 0}}_{\hat \jmath}\,{\Lambda^{\hat\jmath}}_{\bar \imath}
&=& {\Lambda^{\bar 0}}_{\hat \mu}\,{\Lambda^{\hat\mu}}_{\bar \imath} - {\Lambda^{\bar 0}}_{\hat 0}\,{\Lambda^{\hat 0}}_{\bar \imath} \label{vL}\\
&=& - {\Lambda^{\bar 0}}_{\hat 0}\,{\Lambda^{\hat 0}}_{\bar \imath}.
\end{eqnarray}
(The first term in Eq.~(\ref{vL}) vanishes, being equal to ${\delta^{\bar 0}}_{\bar\imath}$.) Hence
\begin{equation}
\e\,{\Lambda^{\bar 0}}_{\hat \jmath}\left({\Lambda^{\hat\jmath}}_{\bar \imath}\,\p{k} - {\Lambda^{\hat\jmath}}_{\bar k}\,\p{\imath}\right)=-\e\,{{\Lambda^{\bar 0}}_{\hat 0}}\left({\Lambda^{\hat 0}}_{\bar \imath}\,\p{k} - {\Lambda^{\hat 0}}_{\bar k}\,\p{\imath}\right).\label{vL2}
\end{equation}
Furthermore, we note that (for example) 
\begin{eqnarray}
p_{\hat \jmath}\, {\Lambda^{\hat \jmath}}_{\bar \imath} &=& p_{\hat \mu}\, {\Lambda^{\hat\mu}}_{\bar \imath} -  p_{\hat 0}\, {\Lambda^{\hat 0}}_{\bar \imath}\\
&=& p_{\bar \imath} + \e\, {\Lambda^{\hat 0}}_{\bar \imath}
\end{eqnarray}
(recall that $p_{\hat 0}=\eta_{\hat 0 \hat\mu}\,p^{\hat \mu} = -p^{\hat 0}$),
so that 
\begin{widetext}
\begin{eqnarray}
{{\Lambda^{\bar 0}}_{\hat 0}\,p_{\hat\jmath}}
\left({\Lambda^{\hat\jmath}}_{\bar \imath}\,\p{k} - {\Lambda^{\hat\jmath}}_{\bar k}\,\p{\imath}\right) &=& {\Lambda^{\bar 0}}_{\hat 0}\left[\left(p_{\bar \imath}\,\p{k} - p_{\bar k}\,\p{\imath}\right)+ \e\left({\Lambda^{\hat 0}}_{\bar \imath}\,\p{k} - {\Lambda^{\hat 0}}_{\bar k}\,\p{\imath}\right)\right] \label{nL2}\\
&=& \e\,{{\Lambda^{\bar 0}}_{\hat 0}}\left({\Lambda^{\hat 0}}_{\bar \imath}\,\p{k} - {\Lambda^{\hat 0}}_{\bar k}\,\p{\imath}\right).\label{nL3}
\end{eqnarray}
(The first term in Eq.~(\ref{nL2}) vanishes because $p_{\bar \imath}=\p{\imath}$ in an orthonormal basis.) Taken together, Eqs.~(\ref{vL2}) and (\ref{nL3}) imply that Eq.~(\ref{EMabc2}) vanishes, proving Eq.~(\ref{EMabc}) as desired.

The ``extra'' term $E_M$ is now reduced to 
\begin{eqnarray}
E_M&=&-{1\over\e}\left(\e\,{\Lambda^{\bar 0}}_{\hat \jmath}+{{\Lambda^{\bar 0}}_{\hat 0}\,p_{\hat\jmath}}\right)
	V^{\hat\jmath}\,f \\
&=&-\e\,{e^{\mu}}_{\bar\mu}\,p^{\bar\mu}\,f\left(\n{j}+\v{j}\right)\left[\c^2{\partial\v{j}\over\partial x^\mu}+{\c\,\v{j}\over \c+1}{\partial\c\over\partial x^\mu}+{\c^3 \over \c+1}\left(\v{k}{\partial\v{j}\over\partial x^\mu}-\v{j}{\partial\v{k}\over\partial x^\mu} \right)\n{k}\right],
\label{EM2}
\end{eqnarray}
where we have used Eqs.~(\ref{boost3}), (\ref{boost0}), and (\ref{comovingMomentum1}-\ref{comovingMomentum3}) to rewrite the leading factor, and Eq.~(\ref{vfactor}) to substitute for $V^{\hat\jmath}$. We denote the quantity in square brackets by $[\dots]$. The last two terms of $[\dots]$ don't contribute to the combination $\n{j}\;[\dots]$, because the contraction of the symmetric combination $\n{j}\n{k}$ with an expression antisymmetric in $j$ and $k$ vanishes; all that remains is 
\begin{equation}
\n{j}\;[\dots] = \c^2\,\n{j}{\partial\v{j}\over\partial x^\mu}+{\c\,\n{j}\v{j}\over \c+1}{\partial\c\over\partial x^\mu}.\label{nV}
\end{equation}
Using the identities $\v{j}\v{j}=(\c-1)(\c+1)/\c^2$ and $\v{j}({\partial\v{j}/\partial x^\mu})=\c^{-3}(\partial\c/\partial x^\mu)$, we find that
\begin{eqnarray}
\v{j}[\dots]&=&\left({1\over\c}+{\c-1\over\c}+{\n{k}\v{k}\over\c+1}\right){\partial\c\over\partial x^\mu}-\c(\c-1)\n{k}{\partial\v{k}\over\partial x^\mu} \\
&=& \left(1+{\n{k}\v{k}\over\c+1}\right){\partial\c\over\partial x^\mu}-(\c^2-\c)\n{k}{\partial\v{k}\over\partial x^\mu}.\label{vV}
\end{eqnarray}
Plugging the sum of Eqs.~(\ref{nV}) and (\ref{vV}) into Eq.~(\ref{EM2}) yields
\begin{equation}
E_M=-\e\,{e^{\mu}}_{\bar\mu}\,p^{\bar\mu}f\left[\c\,\n{i}{\partial\v{i}\over\partial x^\mu}+\left(1+\n{i}\v{i}\right){\partial\gamma\over\partial x^\mu}\right].
\end{equation}
Comparison with Eq.~(\ref{ES}) for $E_S$ shows that the desired cancellation $E_S+E_M=0$ is verified.
\end{widetext}

\begin{acknowledgments}
This work was supported 
by  Scientific Discovery Through
Advanced Computing (SciDAC), a program of the Office of Science of the U.S. Department of Energy (DoE); and by Oak Ridge National Laboratory, managed by UT-Battelle, LLC, for the DoE under contract DE-AC05-00OR22725.
\end{acknowledgments}

\def\aap{Astron. Astrophys.}
\def\apjl{Astrophys. J. Lett.}
\def\apjs{Astrophys. J. Suppl. Ser.}
\def\jcam{J. Comput. Appl. Math.}
\def\mnras{Mon. Not. R. Astron. Soc.}
\def\npa{Nucl. Phys. A}


\end{document}